\documentclass{article}
\usepackage{ijcai26}
\usepackage{times}
\usepackage{soul}
\usepackage{tabularx}
\usepackage{url}
\usepackage[hidelinks]{hyperref}
\usepackage[utf8]{inputenc}
\usepackage[small]{caption}
\usepackage{subfigure}
\usepackage{pifont} 
\usepackage{graphicx}
\usepackage{amsfonts}
\usepackage{amsmath}
\usepackage{amsthm}
\usepackage{array}
\usepackage{multirow}
\usepackage{booktabs}
\usepackage{algorithm}
\usepackage{algorithmic}
\usepackage[switch]{lineno}

\title{TASER: Task-Aware Spectral Energy Refine for Backdoor Suppression in UAV Swarms Decentralized Federated Learning}

\author{Sizhe Huang, Shujie Yang}

\begin{document}

\maketitle

\begin{abstract}
As backdoor attacks in UAV-based decentralized federated learning (DFL) grow increasingly stealthy and sophisticated, existing defenses have likewise escalated in complexity. Yet these defenses, which rely heavily on outlier detection, remain vulnerable to carefully crafted backdoors. In UAV-DFL, the lack of global coordination and limited resources further render outlier-based defenses impractical. Against this backdrop, gradient spectral analysis offers a promising alternative. While prior work primarily leverages low-frequency coefficients for pairwise comparisons, it neglects to analyze the intrinsic spectral characteristics of backdoor gradients. Through empirical analysis of existing stealthy attacks, we reveal a key insight: the more effort attackers invest in mimicking benign behaviors, the more distinct the spectral concentration becomes. Motivated by this, we propose Task-Aware Spectral Energy Refine (TASER)---a decentralized defense framework. To our knowledge, this is the first efficient backdoor defense that utilizes spectral concentration instead of complex outlier detection, enabling mitigation of stealthy attacks by structurally disrupting the backdoor task. To suppress the backdoor task, TASER preserves main-task-relevant frequency coefficients and discards others. We provide theoretical guarantees and demonstrate through experiments that TASER remains effective against stealthy backdoor attacks that bypass outlier-based defenses, achieving attack success rate below 20\% and accuracy loss under 5\%.
\end{abstract}

\section{Introduction}
The integration of decentralized federated learning (DFL) into Unmanned Aerial Vehicle (UAV) swarms has enabled resilient, collaborative model training without relying on a centralized coordinator~\cite{dfl-survey,blockchain_uav}. UAV swarms can collect diverse on-site data and collaboratively train models while preserving data privacy. By leveraging DFL, they eliminate reliance on centralized infrastructure and enable scalable, real-time learning across a distributed network of autonomous agents. However, this integration also introduces new security vulnerabilities, particularly stealthy backdoor attacks, in which malicious participants inject carefully crafted, subtle model perturbations to manipulate predictions on attacker-chosen inputs.

\textbf{Stealthy Backdoor Attacks.}
Modern backdoor attacks have become increasingly stealthy and sophisticated~\cite{lyu2024lurking}. Rather than introducing detectable anomalies, these attacks carefully mimic benign gradient behaviors, thereby evading traditional defenses~\cite{zhang2024backdoor}. In response, defense mechanisms have grown correspondingly complex, often relying on global gradient collection, cross-client comparisons, or computationally expensive outlier detection techniques~\cite{wan2024data}. This escalating arms race imposes substantial overhead, yet increasing defense complexity no longer guarantees effectiveness---particularly in UAV-DFL systems, where computational and communication resources are severely constrained~\cite{javed2024state}. Moreover, the ad hoc deployment of UAV networks leads to a lack of centralized scheduling and secure identity verification, rendering these systems particularly vulnerable to stealthy backdoor attacks~\cite{p2psec}, where compromised nodes inject malicious behaviors into the global model while preserving normal performance on benign tasks~\cite{sun2019can}.

\textbf{Spectral-Based Backdoor Defense.}
Recent studies have validated the effectiveness of frequency-domain representations for model analysis~\cite{wang2020learning,wang2018packing}. In particular, it has been observed that most of the energy in neural network weights concentrates in low-frequency components after applying the Discrete Cosine Transform (DCT)~\cite{secfft}. In the context of federated learning, FreqFed~\cite{fereidooni2023freqfed} has demonstrated the utility of frequency-based filtering for backdoor mitigation. However, FreqFed primarily focuses on pruning high-frequency components to suppress noise for clustering, and does not explicitly investigate the spectral distinctions between benign and backdoored gradients. Moreover, due to the orthogonality of the DCT, the resulting feature space preserves the geometric structure of the original gradients. As such, FreqFed yields no significant improvement over FLAME~\cite{nguyen2022flame}, which performs clustering directly in the gradient space.
\textit{In summary, existing defenses predominantly focus on distributional statistics and overlook whether stealthy backdoor gradients exhibit pronounced energy concentration in mid-to-high frequency bands compared to benign updates.}

\begin{figure}[htbp]
    \centering
    \subfigure[Traditional Backdoor Attack]{
        \includegraphics[width=0.45\linewidth]{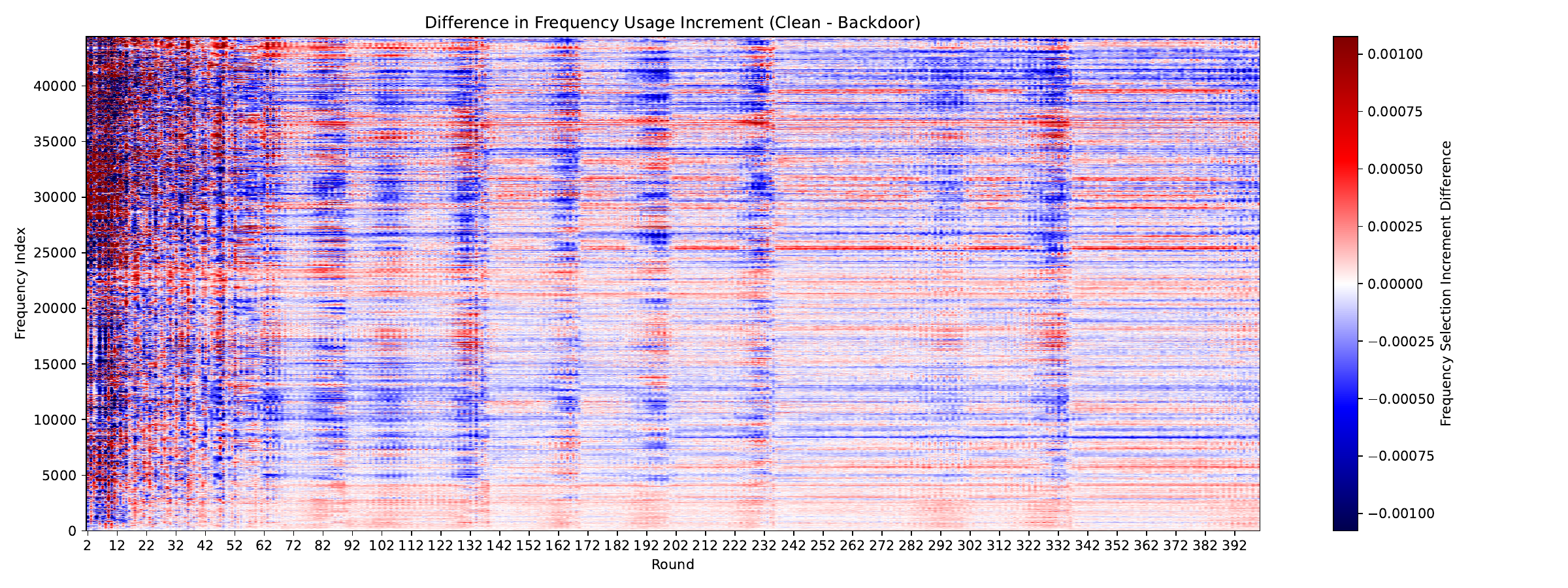}
    }
    \subfigure[Stealthy Backdoor Attack]{
        \includegraphics[width=0.45\linewidth]{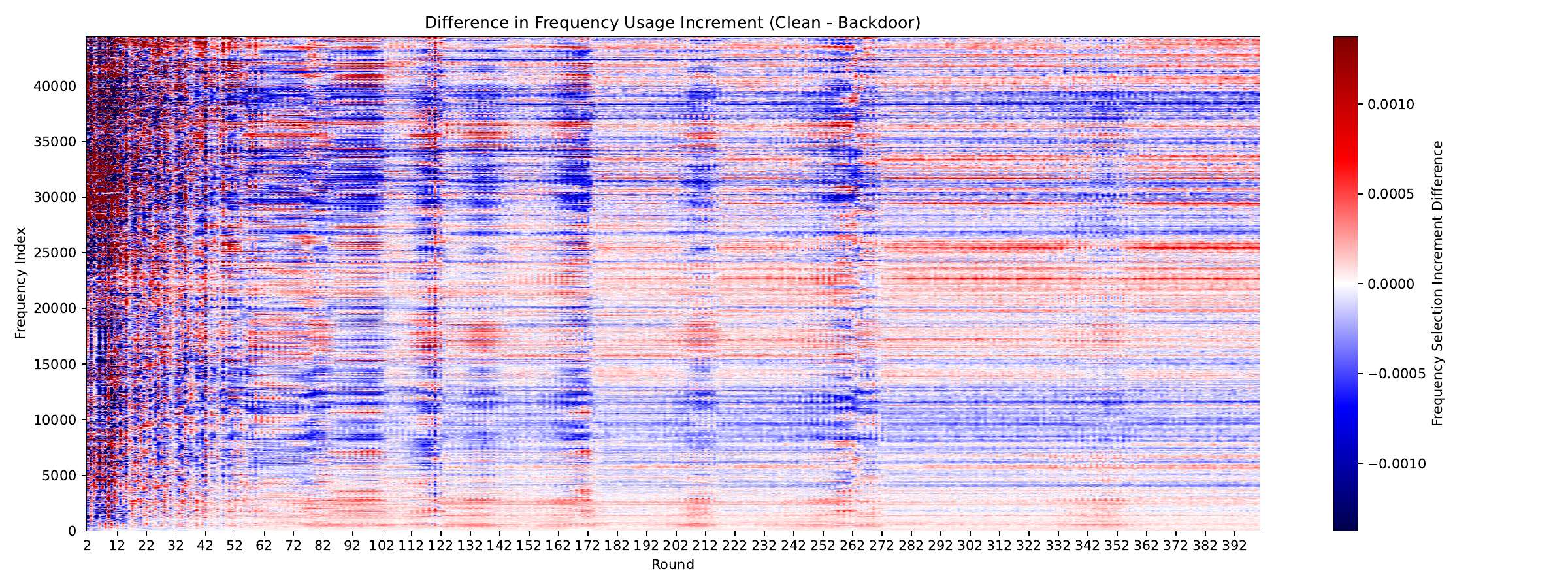}
    }
    \caption{High-energy frequency gradient differences between traditional and stealthy backdoor attacks. We tracked the high-energy frequency indices of 50 clean clients and 50 malicious clients over 400 communication rounds and aggregated their selections into a frequency--time heatmap. The x-axis represents the frequency index, and the y-axis represents the communication round. Blue regions indicate frequencies more frequently selected as high-energy by malicious clients, while red regions indicate those favored by benign clients.}
    \label{fig:freq}
\end{figure}

\textbf{Spectral Insights into Stealthy Backdoor Behavior.}
To visualize these differences, we construct a frequency--time heatmap, where each entry represents the difference in high-energy selection ratio between malicious and benign clients at a particular frequency and communication round. In the heatmap, blue indicates a frequency more frequently selected as high-energy by malicious clients, while red indicates frequencies more favored by benign clients. As shown in Fig.~\ref{fig:freq}, we observe that while stealthy backdoor attacks~\cite{lyu2024lurking} are carefully crafted to align with benign clients in the gradient space to evade detection, \textit{this alignment inadvertently introduces more concentrated energy patterns in the frequency domain, particularly in mid-frequency bands}.

This seemingly counterintuitive phenomenon can be explained as follows. Stealthy adversaries must satisfy two constraints simultaneously: (1) suppressing high-frequency perturbations that would otherwise trigger outlier detection, and (2) mimicking benign updates in low-frequency components to remain hidden. These dual constraints force malicious manipulations into a narrower mid-frequency region, producing a denser and more structured spectral footprint. In essence, the very effort to remain hidden in the parameter space leads to exposure in the frequency space.

In contrast, traditional non-stealthy attacks~\cite{bagdasaryan2020backdoor}, which lack such gradient alignment strategies, often inject arbitrary or exaggerated updates that disperse more broadly across the frequency spectrum. Their spectral signatures are relatively scattered and less consistent over time, reflecting the absence of stealth constraints.

This clear divergence between stealthy and non-stealthy attack patterns---particularly in how spectral energy is distributed---highlights a fundamental vulnerability in stealthy attacks. These findings suggest that analyzing gradients in the frequency domain provides a more discriminative perspective for identifying adversarial behaviors that remain concealed under conventional gradient-space evaluations. This observation motivates the development of frequency-aware defenses that exploit spectral concentration to suppress malicious influence, even when attacks evade norm-based or clustering-based detection strategies.

\textbf{We propose Task-Aware Spectral Energy Refine (TASER), a decentralized defense framework tailored for communication-constrained and coordination-limited UAV-DFL scenarios.}
TASER operates in the frequency domain to exploit the divergent spectral characteristics between benign and backdoor updates. Specifically, each node performs DCT on the gradients computed from multiple local mini-batches, yielding a set of frequency-domain representations. A task-aware scoring metric is then applied to each frequency component, combining approximate Fisher information~\cite{liao2018approximate} and inter-batch consistency via aggregated absolute values to reflect both task sensitivity and directional stability.

Under communication constraints, each node selects the top-$k$ frequency components based on this score and broadcasts their indices to its immediate neighbors. Upon receiving requests, nodes respond with only the requested coefficients. This selective communication protocol significantly reduces bandwidth usage while enabling robust information sharing. The final local update is reconstructed via inverse DCT using the received coefficients.

By retaining only task-relevant frequency components and suppressing anomalous spectral regions typically exploited by backdoor attackers, TASER naturally mitigates adversarial interference. Moreover, its fully decentralized workflow requires no global coordination or intensive computation, making it well-suited for dynamic UAV-based federated systems.

\vspace{0.5em}
\noindent\textbf{Contributions.} Our main contributions are summarized as follows:

\begin{enumerate}
    \item \textbf{Spectral Exposure of Stealthy Backdoor Attacks.}
    To our knowledge, this is the first work revealing that stealthy backdoor attacks exhibit distinctive high-energy concentration patterns in specific frequency bands, offering a novel perspective for defense design.
    
    \item \textbf{Task-Aware Spectral Energy Refine (TASER).}
    We propose TASER, a lightweight, fully decentralized defense mechanism that selects gradient frequency components based on task relevance, effectively suppressing stealthy backdoors under communication constraints.
    
    \item \textbf{Theoretical and Empirical Validation.}
    We provide rigorous convergence analysis and comprehensive experiments demonstrating effectiveness across diverse backdoor scenarios.
\end{enumerate}

\section{Background and Motivation}

\subsection{Challenges in Defending Stealthy Backdoor Attacks}

Stealthy backdoor attacks carefully manipulate the model parameter update process~\cite{gong2022backdoor}, enabling the successful injection of backdoors under the guise of normal update behavior, thereby concealing the attack intent.

These approaches can be categorized into two classes: \textbf{(1) Concealment}, which hides the malicious updates in less sensitive regions of the model, and \textbf{(2) Mimicry}, which forces the malicious behavior to resemble that of benign clients.

\textbf{(1) Concealment:} These attacks inject backdoor perturbations into parameters that are weakly associated with the main task, minimizing interference. Neurotoxin~\cite{zhang2022neurotoxin} uploads poisoned gradients that only modify infrequently updated (``lazy'') parameters, allowing the backdoor to persist across rounds. Similarly, IBA~\cite{nguyen2023iba} targets parameters with low variance across training, incrementally injecting malicious signals that are unlikely to be overwritten by benign updates.

\textbf{(2) Mimicry:} These attacks aim to hide poisoned updates by aligning them with benign contributions, either in the gradient or model parameter space. DEFEAT~\cite{Zhao_2022_CVPR} adds regularization constraints on the intermediate features of the triggered samples to enforce their similarity to clean samples across all layers. PFedBA~\cite{lyu2024lurking} optimizes the backdoor trigger such that its gradient aligns with the direction of the main task in the gradient space, while minimizing the loss difference between the main task and the backdoor task.

By minimizing the divergence between malicious and benign gradient updates, stealthy backdoor attacks pose a significant challenge to existing defense mechanisms that rely on detecting malicious update gradients.

\subsection{Limitations of Existing Backdoor Defenses in UAV-DFL}

Decentralized federated learning (DFL) over UAV networks imposes several critical constraints on backdoor defense design: (1) partial observability, as each UAV can only communicate with a limited set of neighbors; (2) constrained computation and energy resources, which limit the feasibility of high-complexity algorithms; and (3) dynamic and unstable network topology, which hinders long-term coordination or global trust modeling.

\textbf{Robust aggregation approaches require global visibility} to collect client updates and apply global aggregation rules, such as RFA~\cite{pillutla2022robust} and RLR~\cite{ozdayi2021defending}. RFA updates the global model by the geometric median of gradients, while RLR adjusts the dimension-wise learning rate to corrupt the dimensions updated by the backdoor. However, in UAV-DFL, each client can only access a small subset of updates from its directly connected neighbors in each round, making it difficult to identify benign gradients.

\textbf{Detection-based defenses require comparison of model updates}, such as FreqFed~\cite{fereidooni2023freqfed} and Multi-Metrics~\cite{huang2023multi}. FreqFed utilizes the cosine similarity between the low-frequency components of gradients as the clustering metric in HDBSCAN. Multi-Metrics utilizes multiple metrics with dynamic weighting to detect backdoor attacks. However, these methods typically incur $O(n^2d)$ computational complexity for $n$ clients and $d$-dimensional models, which is prohibitive for UAV platforms with constrained computational and energy resources.

\textbf{Credit-based defenses assume a stable network topology for long-term reputation tracking}, such as Baffle~\cite{andreina2021baffle} and FLDetector~\cite{zhang2022fldetector}. FLDetector detects malicious clients by checking the consistency of their model updates across rounds. Baffle detects poisoned updates via client feedback and per-class performance evaluation. However, UAV networks are often highly dynamic and ad hoc~\cite{gupta2015survey}, with rapidly changing communication links and participation sets. As a result, constructing reliable trust models becomes exceedingly difficult, and reputation-based defenses cannot be effectively applied.

Collectively, these limitations reveal a fundamental mismatch between existing defense designs and the decentralized, resource-constrained, and dynamically connected nature of UAV-based federated networks.

While most backdoor defenses are inapplicable in decentralized UAV networks, \textbf{model purification-based approaches offer an alternative direction}. These methods typically operate by pruning backdoor-sensitive neurons~\cite{wu2021adversarial} or injecting noise into intermediate layers~\cite{sun2019can}, thereby suppressing hidden triggers. However, their computation and inference cost is prohibitive for UAVs with limited resources. In addition, injecting noise can severely degrade benign performance if it is not guided by task relevance.

\subsection{Motivation}

Stealthy backdoor attacks are known to exhibit greater sensitivity to model perturbations than benign tasks~\cite{wu2021adversarial}, meaning that even small modifications to the model may significantly disrupt their effectiveness. Complementing this, our empirical observations reveal that such attacks tend to concentrate their influence in a narrow range of frequency components when gradients are transformed via DCT.

This insight motivates a frequency-aware defense strategy: by selectively discarding frequency components that are both (i) less relevant to the main task and (ii) more likely to carry backdoor-related perturbations, it becomes possible to suppress hidden triggers while retaining benign performance. To this end, we propose a lightweight decentralized mechanism that estimates task relevance in the frequency domain and filters model updates accordingly.
\section{Preliminaries}

\subsection{Discrete Cosine Transform for Gradient Analysis}

The Discrete Cosine Transform (DCT) is a frequency-domain transformation that decomposes a signal into a weighted sum of cosine functions at different frequencies. For a gradient vector $\mathbf{g} \in \mathbb{R}^n$, the DCT coefficients are computed as:
\begin{equation}
    G_k = \sum_{i=0}^{n-1} g_i \cos\left[\frac{\pi}{n}\left(i + \frac{1}{2}\right)k\right], \quad k = 0, 1, \ldots, n-1
\end{equation}
where $G_k$ denotes the $k$-th frequency coefficient.

\textbf{Frequency Components.} In the DCT representation, \textit{low-frequency components} (small $k$) capture smooth, global patterns in the gradient that typically correspond to the overall learning direction. \textit{High-frequency components} (large $k$) encode rapid local variations, which often include noise or fine-grained perturbations. \textit{Mid-frequency components} lie between these extremes and capture intermediate structural information.

\textbf{High-Energy Components.} The energy of a frequency component is defined as $|G_k|^2$. \textit{High-energy components} refer to frequency indices where $|G_k|^2$ is large relative to other components, indicating significant contribution to the overall gradient structure. In neural network gradients, energy is typically concentrated in low-frequency components due to the inherent smoothness of loss landscapes~\cite{secfft}.

\textbf{Why DCT for Backdoor Defense?} DCT is particularly suitable for gradient analysis in backdoor defense for three reasons:
\begin{enumerate}
    \item \textbf{Energy Compaction}: DCT concentrates most signal energy into a few coefficients, making it easier to identify anomalous energy distributions introduced by backdoor attacks.
    \item \textbf{Structure Preservation}: Due to its orthogonality, DCT preserves the geometric relationships in the original gradient space while revealing hidden spectral patterns.
    \item \textbf{Computational Efficiency}: The Fast Cosine Transform enables $O(n \log n)$ computation, making it practical for resource-constrained UAV systems.
\end{enumerate}

\textbf{Intuition Behind Spectral Exposure.} Frequencies with larger accumulated energy tend to correspond to parameters that are more sensitive to task-specific loss variations, and thus encode stronger task relevance. When stealthy attackers attempt to align their malicious gradients with benign ones in the parameter space, they inadvertently create distinctive patterns in the frequency domain. Specifically, suppressing detectable high-frequency anomalies while mimicking low-frequency benign behavior forces the backdoor signal into mid-frequency bands, resulting in anomalous energy concentration that can be detected and filtered.

\section{System Overview}

\subsection{System Setting and Threat Model}

We consider a decentralized federated learning (DFL) scenario deployed across a swarm of UAVs.

\begin{figure*}[t]
    \centering
    \includegraphics[width=0.9\linewidth]{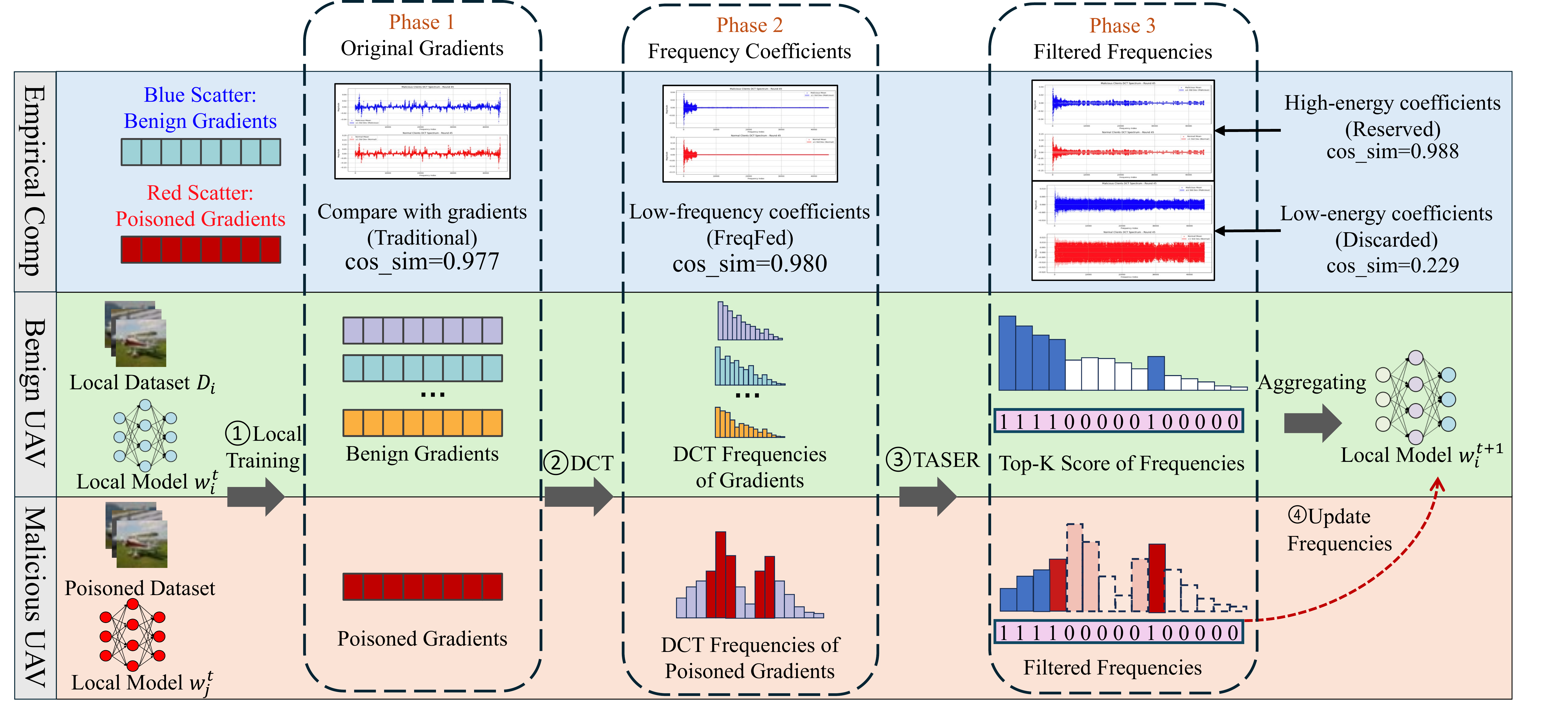} 
    \caption{Workflow of TASER}
    \label{fig:workflow}
\end{figure*}

As shown in Fig.~\ref{fig:workflow}, let the total number of UAV clients be $N$, and each UAV $i$ $(1 \leq i \leq N)$ has a local dataset $\mathcal{D}_i$ consisting of clean training samples.

Each node performs training on its local data and engages in communication only with its immediate neighbors over an ad-hoc wireless network.

To simplify the communication workflow, global training is conducted in synchronous communication rounds indexed by $t = 1, 2, \dots, T$. In each round, node $i$ updates its local model parameters $\theta_i^t$ by performing stochastic gradient descent (SGD) over its dataset $\mathcal{D}_i$, resulting in a model update $\Delta \theta_i^t$. Let $g_i^t = \Delta \theta_i^t$ denote the gradient or model delta computed by client $i$ at round $t$.

Due to limited bandwidth constraints on UAVs, transmitting full model updates is inefficient. Hence, each node should adopt communication-efficient strategies when sharing updates.

\subsection{Threat Model}

We assume a stealthy backdoor threat model. A subset of UAV clients $\mathcal{A} \subset \{1, \dots, N\}$ are compromised and attempt to implant a backdoor task into the global model.

These adversarial nodes create malicious updates $g_{i}^t$ for $i \in \mathcal{A}$ that preserve the accuracy of the main task while encoding the behavior of targeted misclassifications on triggered inputs.

The system lacks global coordination or a centralized verifier, and any node may potentially be adversarial. Furthermore, attackers may exploit gradient manipulation strategies such as scaling, clipping, or directional alignment to evade detection.

\subsection{Workflow}

In each communication round $t$, every node $i$ performs the following steps:

\begin{enumerate}
\item \textbf{Local Update:}
Node $i$ trains its local model on dataset $\mathcal{D}_i$ and obtains the gradient $g_i^t$.

\item \textbf{Frequency Transformation:}
For each mini-batch gradient $g_{i,b}^t$, node $i$ applies DCT to obtain $\hat{g}_{i,b}^t = \text{DCT}(g_{i,b}^t)$. The set of transformed gradients is denoted as $\{\hat{g}_{i,b}^t\}_{b=1}^B$.

\item \textbf{Task-Aware Scoring:}
Node $i$ aggregates information across mini-batches to compute a frequency-wise score: 
\[
\text{score}_i^t = \sum_{b=1}^B \left(\hat{g}_{i,b}^t\right)^2 + \left|\sum_{b=1}^B \hat{g}_{i,b}^t\right|,
\]
where the first term serves as an approximation to the diagonal of the Fisher information matrix in the frequency domain, capturing the relative importance of each frequency component to the training objective, and the second term captures cumulative frequency energy.

\item \textbf{Top-$k$ Selection:}
Based on the magnitude of $\text{score}_i^t$, node $i$ selects the top-$k$ frequency indices $\mathcal{K}_i^t$ to retain, where $k$ is determined by communication constraints.

\item \textbf{Frequency Request:}
Node $i$ broadcasts a request message containing its index set $\mathcal{K}_i^t$ to its neighbors and listens for incoming requests.

\item \textbf{Selective Communication and Aggregation:}
Node $i$ transmits to each requesting neighbor the coefficients they asked for from its own aggregated DCT spectrum, and in turn receives the requested coefficients from neighbors. It then aggregates received frequency components and reconstructs the final update $\tilde{g}_i^t$ via inverse DCT.

\item \textbf{Model Update:}
Node $i$ updates its local model using the reconstructed aggregate $\tilde{g}_i^t$.
\end{enumerate}

This decentralized, task-driven workflow enables efficient information exchange under bandwidth constraints while suppressing anomalous frequency components injected by backdoor clients.

\section{Methodology}

In this section, we present TASER to defend against backdoor attacks under UAV-DFL scenarios.

\subsection{Frequency Transformation}

To reveal the energy structure of gradient updates, we apply the Discrete Cosine Transform (DCT) to project them into the frequency domain~\cite{secfft}.

DCT is known for its \emph{energy compaction} property---benign updates typically concentrate most of their energy in low-frequency components, while adversarial perturbations or abrupt changes are more likely to activate mid-to-high frequency bands.

For each client $i$ at communication round $t$, we denote the mini-batch gradients as $\{g_{i,b}^t\}_{b=1}^B$, where each $g_{i,b}^t \in \mathbb{R}^d$ represents the flattened gradient vector computed from batch $b$ of the local clean dataset $\mathcal{D}_i$. Each gradient is projected into the frequency domain as:
\[
\hat{g}_{i,b}^t = \text{DCT}(g_{i,b}^t),
\]
where DCT is applied using an orthonormal basis.

Specifically, for a gradient vector $g \in \mathbb{R}^d$, its DCT (type-II) is defined as:
\[
\hat{g}[k] = \sqrt{\frac{2}{d}} \sum_{n=0}^{d-1} g[n] \cos\left( \frac{\pi (2n + 1)k}{2d} \right), \quad \text{for } 0 \leq k < d.
\]
For $k = 0$, we use the normalization factor $\sqrt{1/d}$ to ensure orthogonality:
\[
\hat{g}[0] = \sqrt{\frac{1}{d}} \sum_{n=0}^{d-1} g[n].
\]

The resulting set $\{\hat{g}_{i,b}^t\}_{b=1}^B$ serves as the foundation for our subsequent scoring and filtering mechanism.

\subsection{Task-Aware Frequency Scoring}

After transforming gradient updates to the frequency domain, we aim to quantify the importance of each frequency component in a manner that reflects both task relevance and robustness to adversarial manipulation. Our design is motivated by the following observations:

\begin{itemize}
    \item \textbf{Gradient Energy Distribution Reflects Task Sensitivity.} Frequency components with consistently high energy across mini-batches are more likely to encode stable task-relevant information.
    \item \textbf{Directional Consistency Implies Optimization Alignment.} Frequencies whose directions remain stable across updates suggest convergence toward a shared local objective and are less likely to originate from stochastic noise.
\end{itemize}

Based on these principles, we define a frequency-wise scoring function composed of two terms:

\[
\text{score}_i^t[k] = \alpha \cdot E_i^t[k] + (1 - \alpha) \cdot D_i^t[k],
\]
where
\[
E_i^t[k] = \sum_{b=1}^B \left( \hat{g}_{i,b}^t[k] \right)^2, \quad D_i^t[k] = \left| \sum_{b=1}^B \hat{g}_{i,b}^t[k] \right|.
\]

The first term $E_i^t[k]$ captures the accumulated gradient energy in the $k$-th frequency component across local mini-batches. Due to the orthogonality of DCT, this value serves as a local proxy for task sensitivity in the $k$-th frequency direction, consistent with the intuition behind diagonal Fisher information:
\[
E_i^t[k] \approx \mathbb{E}_b \left[ \left( \frac{\partial \ell}{\partial \hat{g}_{i,b}^t[k]} \right)^2 \right].
\]
It reflects the degree to which this frequency direction consistently participates in optimization dynamics. Frequencies with larger accumulated energy tend to correspond to parameters more sensitive to task-specific loss variations and thus encode stronger task relevance.

The second term $D_i^t[k]$ captures directional agreement of the gradients. Frequencies with aligned update directions are likely driven by task-level learning dynamics, whereas adversarial components typically vary in direction to evade detection or align selectively.

By combining $E_i^t[k]$ and $D_i^t[k]$, the scoring function favors components that are both stable and semantically meaningful for the main task. The hyperparameter $\alpha \in [0, 1]$ balances the influence between frequency energy and directional coherence. This enables each node to prioritize trustworthy update elements under local view and communication constraints---without relying on pairwise comparison or global consensus.

The final score vector $\text{score}_i^t \in \mathbb{R}^d$ serves as the input for the subsequent Top-$k$ selection step.

\subsection{Top-$k$ Frequency Filtering}

Given the task-aware score vector $\text{score}_i^t \in \mathbb{R}^d$ computed for each frequency component, node $i$ performs a sparsification step to select the most informative and trustworthy update elements under communication constraints.

\paragraph{Dual-purpose filtering.}
The top-$k$ frequency selection plays two roles:

\begin{itemize}
    \item \textbf{Communication Efficiency.}
    Only the $k$ most informative frequency components are retained and exchanged, significantly reducing transmission volume per node.

    \item \textbf{Backdoor Filtering.}
    Malicious updates are naturally assigned lower scores and filtered out in this process.
\end{itemize}

Specifically, the node retains only the indices of the top-$k$ scoring frequencies:

\[
\mathcal{K}_i^t = \text{TopK}\left( \text{score}_i^t, \, k \right),
\]

where $k$ is determined adaptively according to both system-level and security-level constraints. In particular, we define:

\[
k = \min \left\{ k_{\text{bw}}, \, k_{\text{adv}} \right\},
\]

where $k_{\text{bw}}$ is the maximum number of coefficients that can be transmitted under current communication bandwidth and noise level, and $k_{\text{adv}}$ is a security-driven lower bound determined by empirical analysis of backdoor spread across frequencies.

\paragraph{Bandwidth-constrained upper bound.}
Let each node transmit $k$ floating-point values of size $b$ bits per communication round, over a noisy wireless link with signal-to-noise ratio (SNR) $\gamma$ and channel bandwidth $W$. Then, the Shannon capacity gives an upper limit on data rate per node:

\[
R = W \log_2(1 + \gamma).
\]

Thus, the maximum permissible $k$ under communication constraints is:

\[
k_{\text{bw}} = \left\lfloor \frac{R \cdot \Delta t}{b} \right\rfloor,
\]

where $\Delta t$ is the duration of each communication slot.

\paragraph{Security-driven lower bound.}
To ensure sufficient pruning of adversarial signals, $k$ must also be larger than a minimum threshold $k_{\text{adv}}$ that ensures the exclusion of low-scoring, potentially malicious frequencies. This threshold can be determined empirically by observing the frequency rank distributions of known attacks.

Each node then broadcasts its selected frequency indices $\mathcal{K}_i^t$ to neighbors, receives matching coefficients in response, and reconstructs the aggregated update via inverse DCT.

\section{Experiment}

To evaluate the effectiveness and robustness of our proposed method, we conduct experiments on two standard image classification benchmarks: EMNIST~\cite{cohen2017emnist} and CIFAR-10~\cite{cifar10}. EMNIST is a dataset of handwritten alphanumeric characters comprising 62 classes, for which we adopt the LeNet~\cite{lecun2002lenet} architecture. CIFAR-10 consists of 10 categories of natural color images with higher visual complexity, for which we employ a deeper VGG-9~\cite{simonyan2014vgg} network. These two datasets provide complementary testbeds to assess defense performance under varying input domains, model depths, and task complexities.

We compare our method against the following representative federated backdoor defenses: \textbf{Weak-DP}~\cite{sun2019can}, \textbf{Multi-Metrics}~\cite{huang2023multi}, \textbf{Krum \& Multi-Krum}~\cite{blanchard2017machine}, \textbf{RFA}~\cite{pillutla2022robust}, and \textbf{Freqfed}\cite{fereidooni2023freqfed}. We consider both \textbf{black-box stealthy backdoor attacks}~\cite{youcanreally}, which leverage rare edge-case samples to avoid dilution during aggregation, and \textbf{white-box stealthy backdoor attacks} (PFedBA)~\cite{lyu2024lurking}, which evade detection by aligning the injected update direction with clean task gradients.

All experiments are conducted using PyTorch 2.5.1 and Python 3.12 on a virtualized GPU with 48\,GB memory, a 20-core Intel Xeon Platinum 8470Q processor, and 90\,GB RAM. We simulate a decentralized federated learning environment with 200 clients, where 10 clients are randomly selected per round over 100 communication rounds. Among all clients, 20\% (40 clients) are malicious, each holding 30\% poisoned data embedded with a fixed trigger pattern.

\subsection{Impact of Frequency Selection Ratio}

We conduct an ablation study by varying the Top-$k$ selection ratio from 5\% to 100\% while keeping the Poisoned Model Rate (PMR) at 20\%. Experiments are performed on CIFAR-10 under white-box stealthy backdoor attacks (PFedBA).

\begin{figure}[htbp]
    \centering
    \subfigure[Main Task Accuracy]{
        \includegraphics[width=0.45\linewidth]{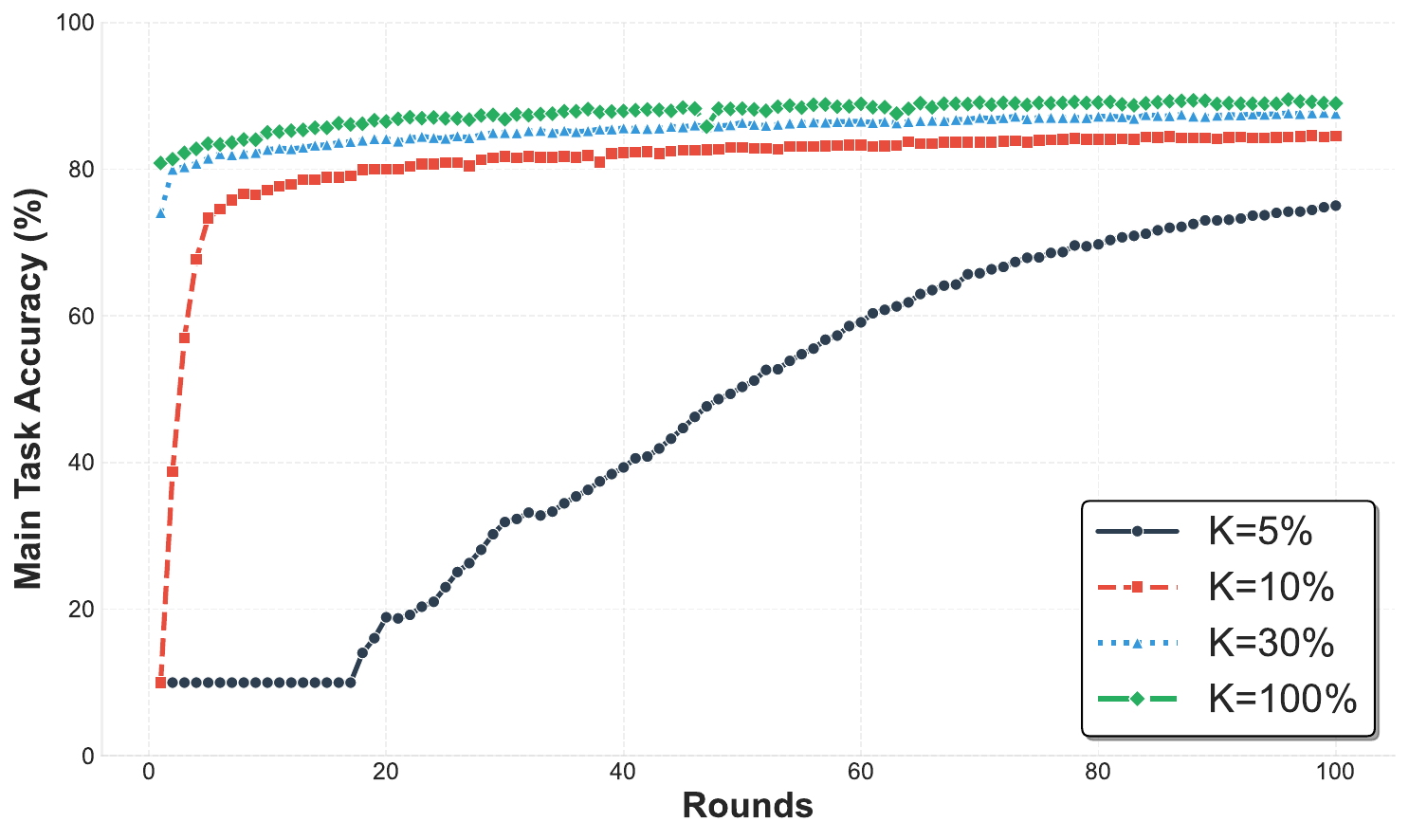}
    }
    \hfill
    \subfigure[Attack Success Rate]{
        \includegraphics[width=0.45\linewidth]{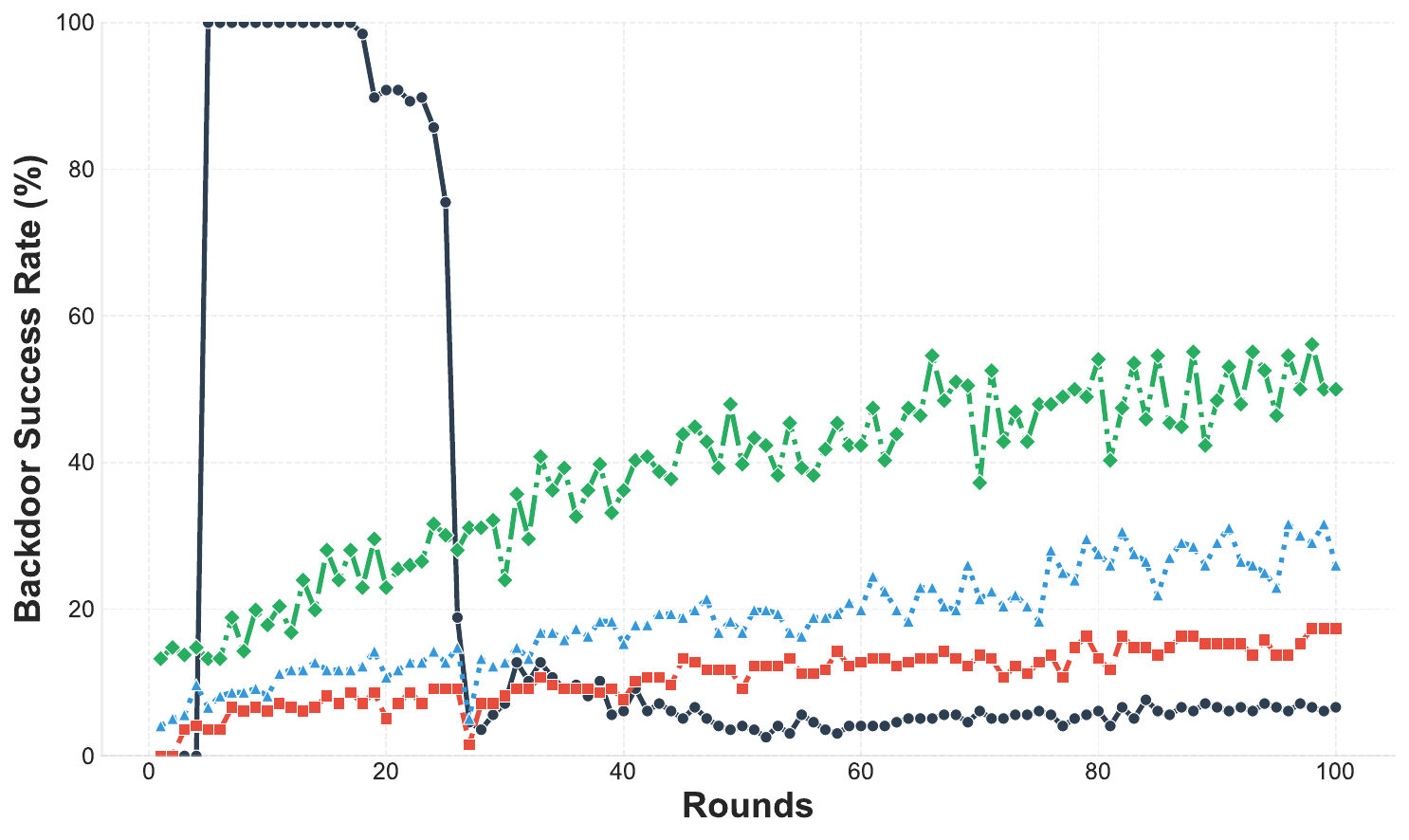}
    }
    \caption{Comparison of main task accuracy and backdoor success rate under different frequency selection ratios.}
    \label{fig:main_vs_backdoor}
\end{figure}

As shown in Fig.~\ref{fig:main_vs_backdoor}, reducing $k$ degrades both main-task accuracy and backdoor success rate, but the decline in backdoor performance is significantly steeper. At $k=30\%$, the retained frequency band still includes regions where stealthy attacks concentrate their manipulations, allowing moderate backdoor persistence. At $k=10\%$, our method achieves sharp suppression of the backdoor effect (ASR $\approx$ 15\%) while maintaining main-task accuracy within 5\% of the baseline. However, $k=5\%$ causes severe performance degradation, indicating over-aggressive pruning.

\textbf{Practical Guidance.} We recommend $k \in [10\%, 20\%]$ as a default configuration: $k=10\%$ for high-adversarial scenarios, and $k=20\%$ for benign-dominated environments.

\subsection{Comparison with Baseline Defenses}

We evaluate Attack Success Rate (ASR) and Main Task Accuracy (MTA) for all defense methods on both datasets under black-box and white-box attacks.

\begin{figure}[htbp]
    \centering
    \subfigure[Main Task Accuracy]{
        \includegraphics[width=0.45\linewidth]{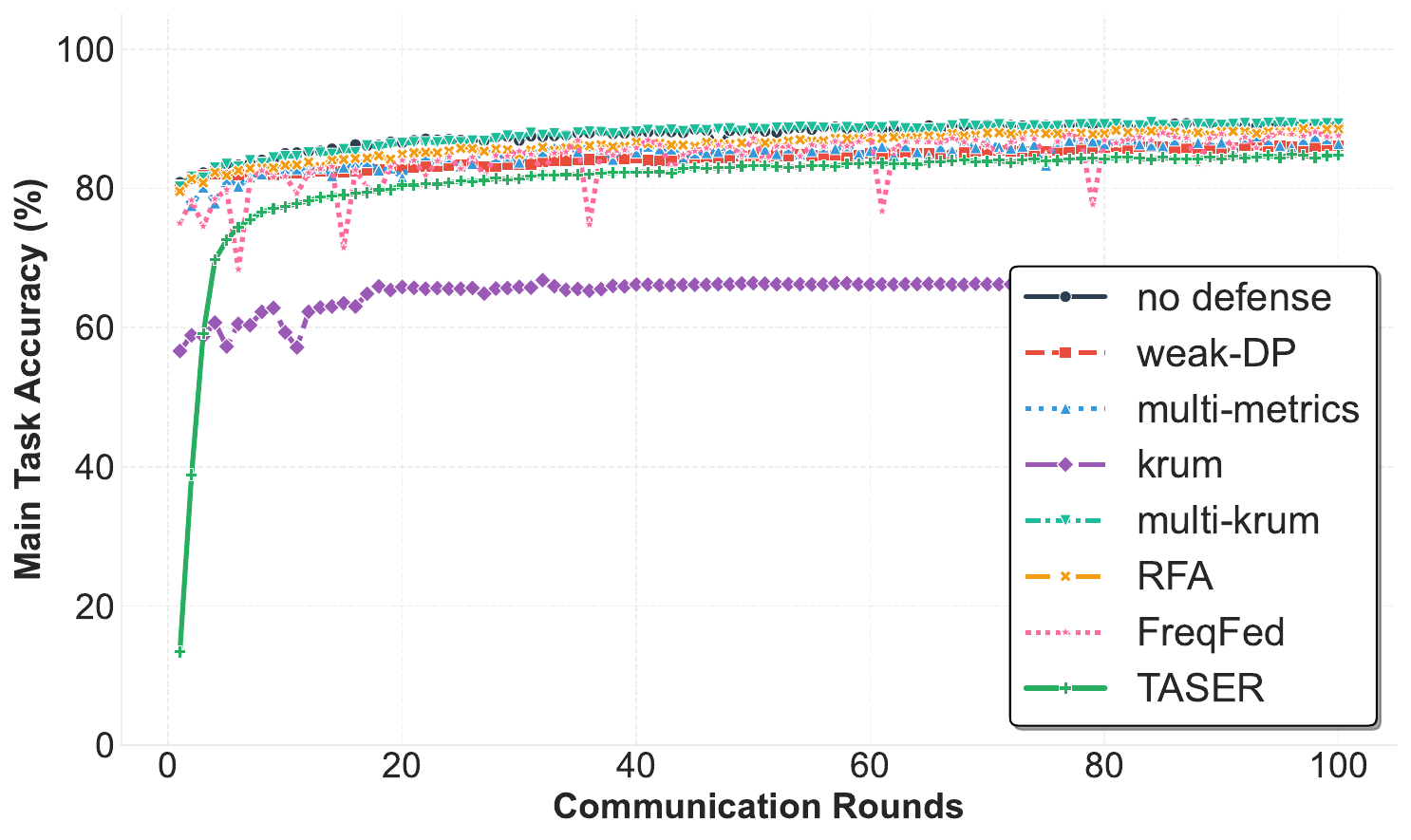}
    }
    \hfill
    \subfigure[Attack Success Rate]{
        \includegraphics[width=0.45\linewidth]{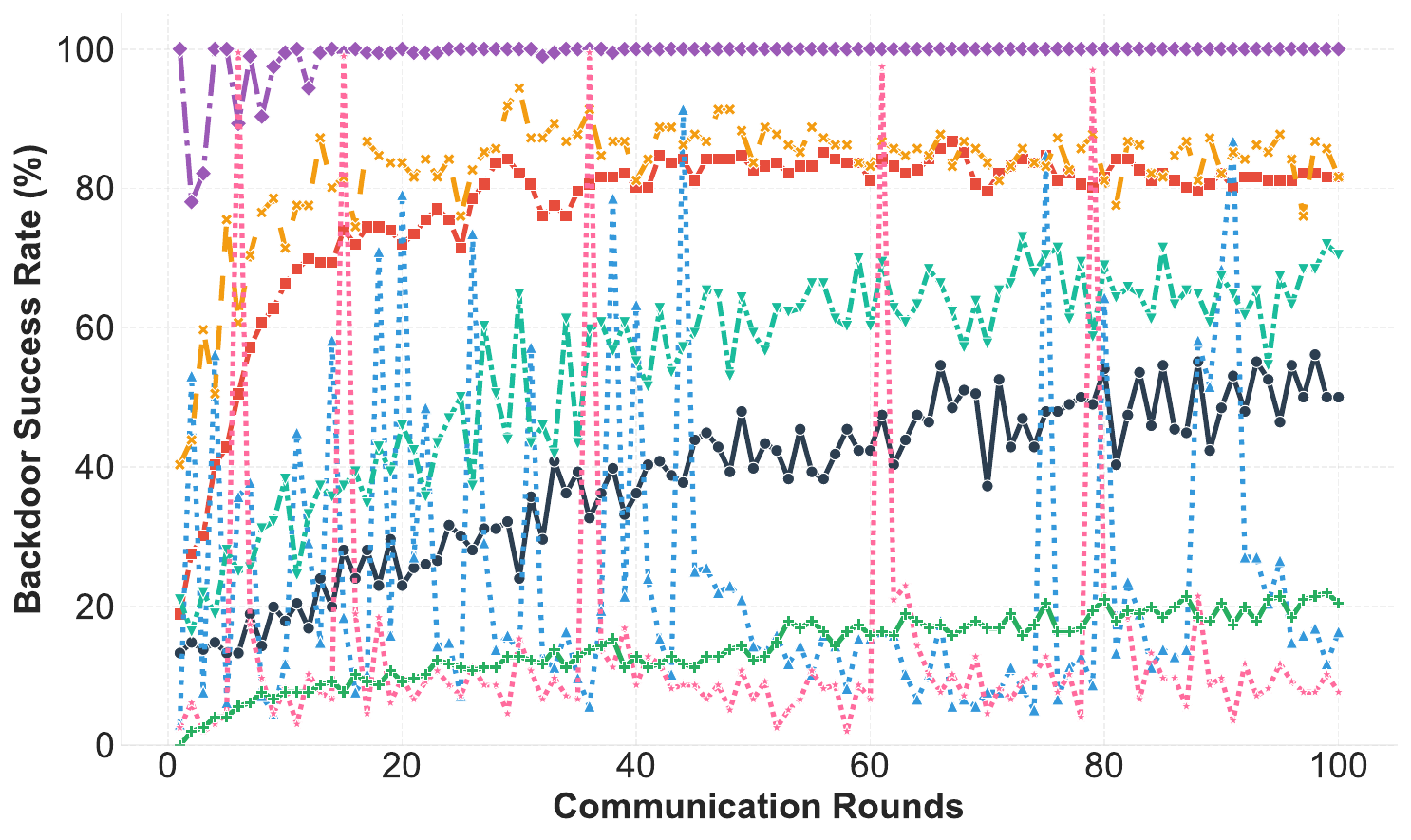}
    }
    \caption{CIFAR-10 against black-box stealthy backdoor attacks.}
    \label{fig:cifar_black_main_vs_backdoor}
\end{figure}

\begin{figure}[htbp]
    \centering
    \subfigure[Main Task Accuracy]{
        \includegraphics[width=0.45\linewidth]{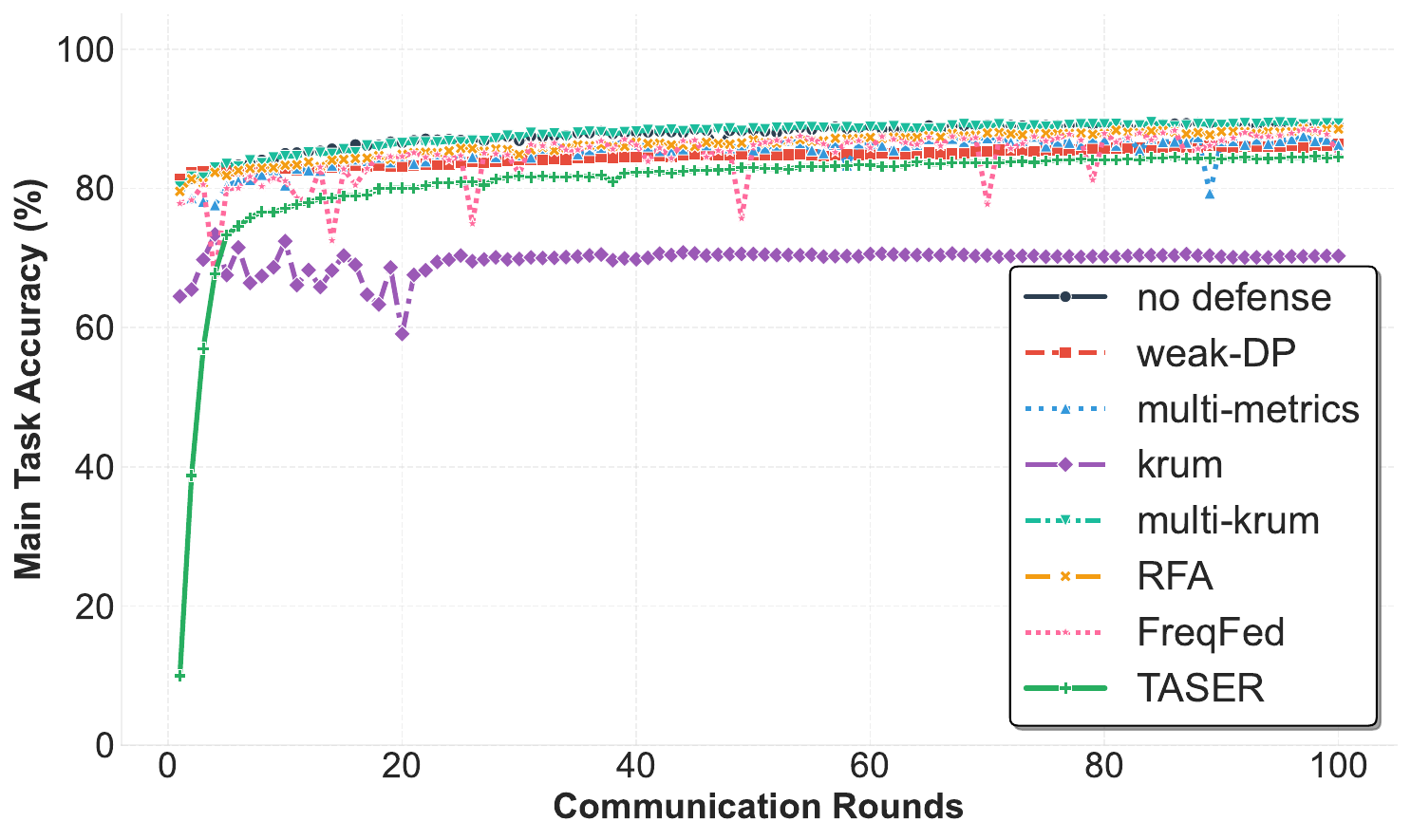}
    }
    \hfill
    \subfigure[Attack Success Rate]{
        \includegraphics[width=0.45\linewidth]{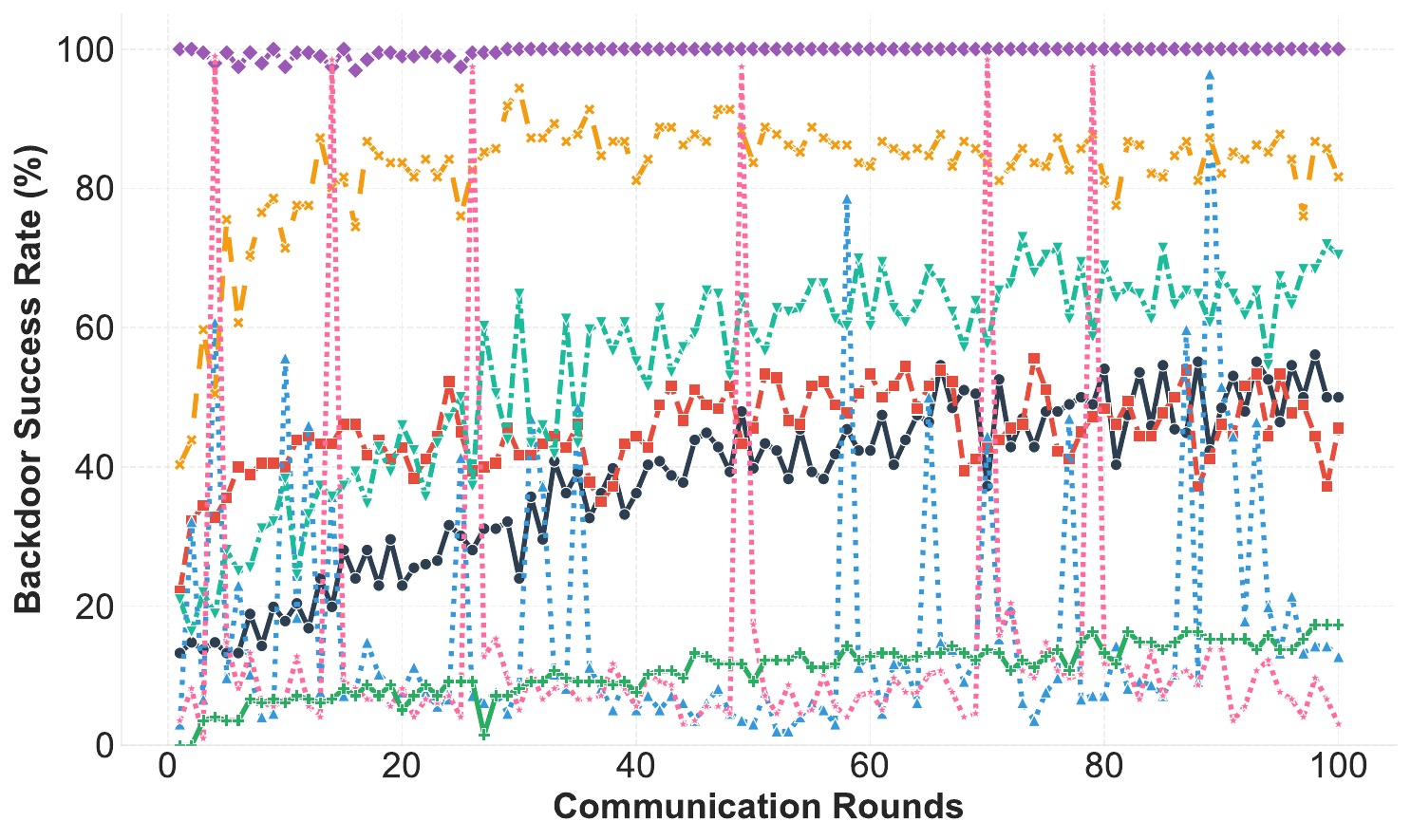}
    }
    \caption{CIFAR-10 against white-box stealthy backdoor attacks.}
    \label{fig:cifar_cos_main_vs_backdoor}
\end{figure}

\begin{figure}[htbp]
    \centering
    \subfigure[Main Task Accuracy]{
        \includegraphics[width=0.45\linewidth]{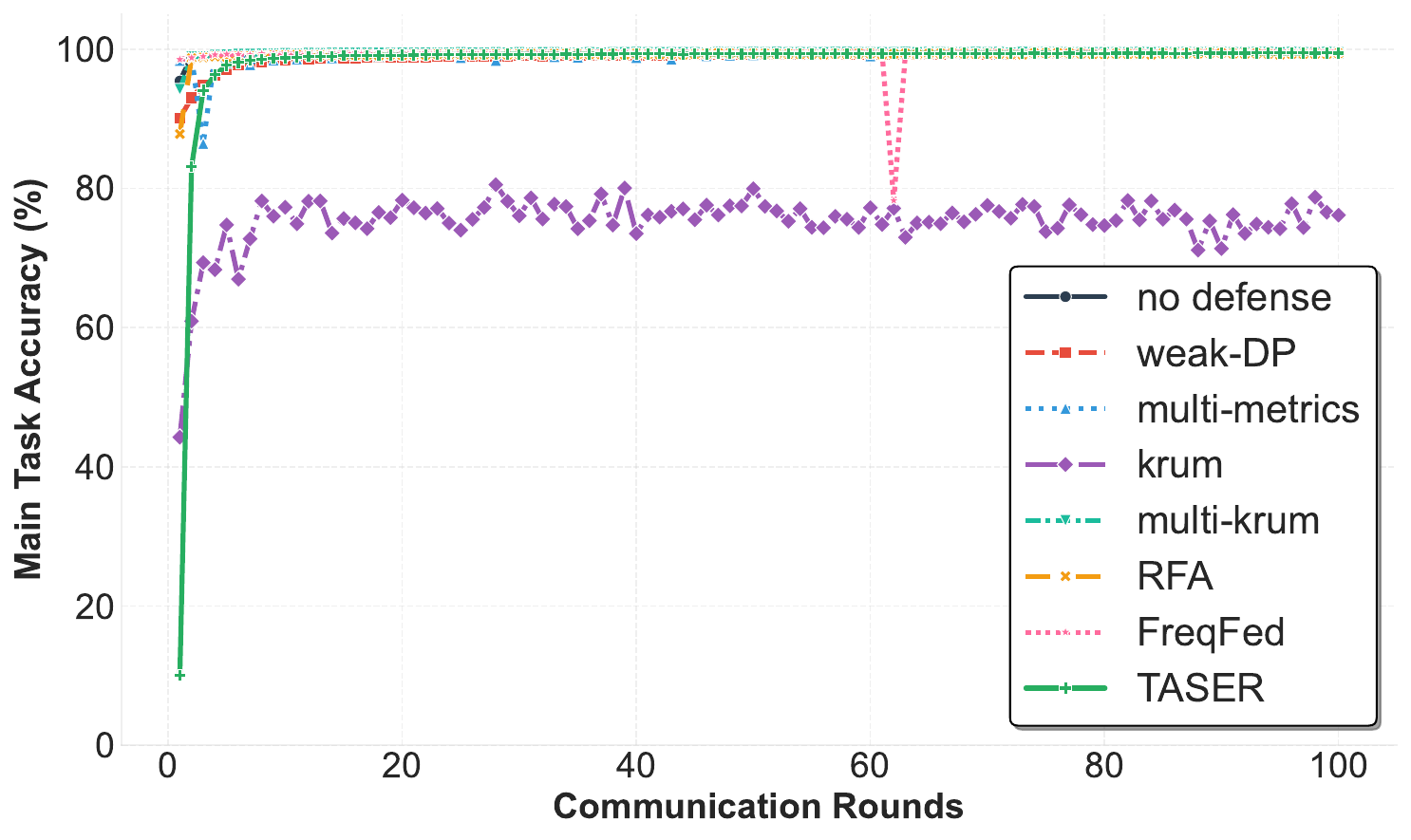}
    }
    \hfill
    \subfigure[Attack Success Rate]{
        \includegraphics[width=0.45\linewidth]{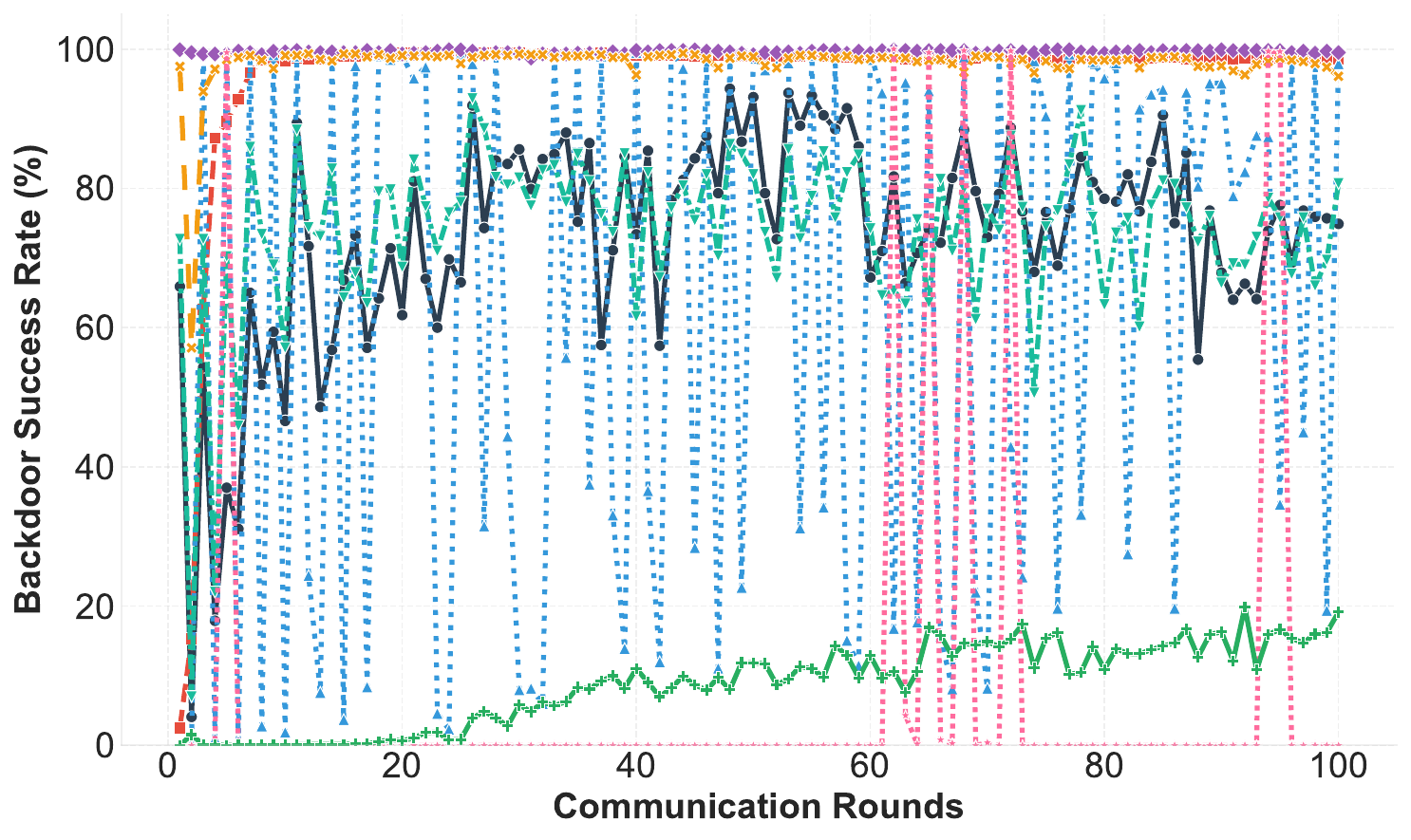}
    }
    \caption{EMNIST against black-box stealthy backdoor attacks.}
    \label{fig:emnist_blackbox_main_vs_backdoor}
\end{figure}

\begin{figure}[htbp]
    \centering
    \subfigure[Main Task Accuracy]{
        \includegraphics[width=0.45\linewidth]{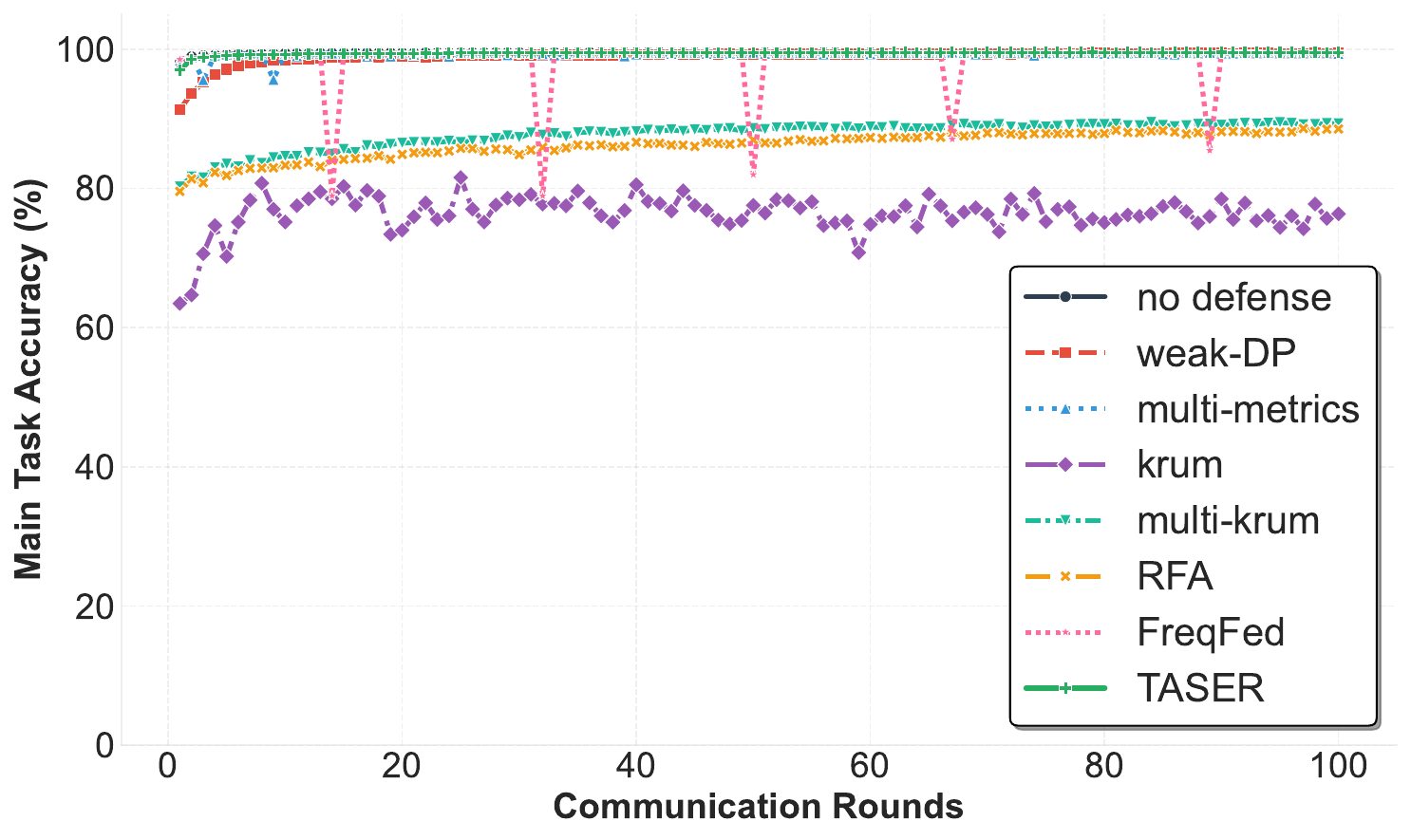}
    }
    \hfill
    \subfigure[Attack Success Rate]{
        \includegraphics[width=0.45\linewidth]{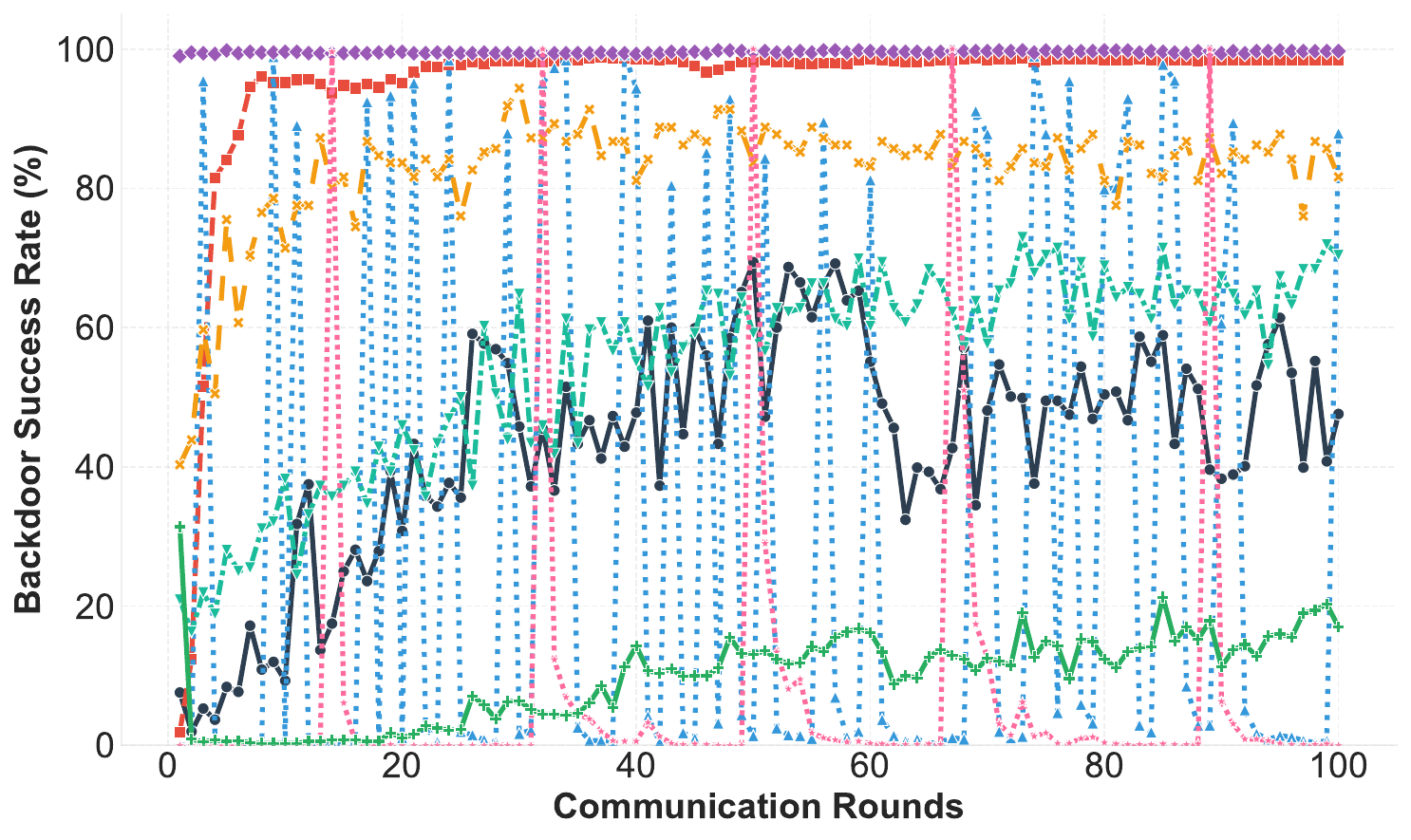}
    }
    \caption{EMNIST against white-box stealthy backdoor attacks.}
    \label{fig:emnist_cos_main_vs_backdoor}
\end{figure}

\textbf{Analysis of Baseline Defenses.}
As shown in Figs.~\ref{fig:cifar_black_main_vs_backdoor}--\ref{fig:emnist_cos_main_vs_backdoor}, robust aggregation methods (Krum, Multi-Krum, RFA) are largely ineffective against stealthy backdoor attacks. By aligning malicious updates with benign gradients, attackers bypass outlier filtering and maintain high ASR. This reliance on distance-based heuristics proves fundamentally insufficient against well-camouflaged adversarial behavior, sometimes resulting in higher ASR than the no-defense setting due to misidentifying malicious clients as benign.

Weak-DP provides virtually no protection despite minimal overhead. Multi-Metrics shows good performance on CIFAR-10 but suffers from non-deterministic failures---sudden ASR spikes occur when malicious clients evade detection. On EMNIST, Multi-Metrics exhibits significantly increased misclassifications due to elevated intra-class similarity in fine-grained tasks.

For FreqFed~\cite{fereidooni2023freqfed}, the frequency-based method exhibits much better performance than multi-metrics. However, it still faces the issue common to clustering-based backdoor detection methods: once a cluster containing a large number of attackers is selected, the backdoor task accuracy will sharply increase in the following round.

\textbf{Performance of TASER.}
Our method selectively preserves task-critical frequency components while suppressing backdoor-related frequencies. On CIFAR-10, main-task accuracy drops by only 5\% compared to no defense; on EMNIST, performance remains virtually unaffected. More importantly, TASER consistently suppresses ASR below 20\% across both datasets, significantly outperforming all baselines. Unlike metric-based defenses, our method maintains stable performance without abrupt vulnerability spikes, confirming the effectiveness of task-driven frequency filtering.
\section{Conclusion}

We introduce TASER, a novel defense framework designed to mitigate stealthy backdoor attacks in decentralized federated learning, particularly in UAV-based scenarios lacking global coordination and identity stability. Instead of relying on outlier detection in the gradient space, TASER leverages the frequency-domain structure of model updates and selectively retains main-task-relevant components, thereby structurally suppressing backdoor effects. Theoretical analysis and empirical results across multiple datasets and threat models demonstrate that TASER achieves strong resistance to both white-box and black-box attacks while maintaining competitive main-task performance. Our findings highlight the potential of task-aware spectral filtering as a lightweight and effective alternative to distribution-based defenses.

\bibliographystyle{named}
\bibliography{ijcai26}
\clearpage

\end{document}